\documentclass[aps,showpacs,superscrpitaddress]{revtex4}
\usepackage{graphicx}
\usepackage{setspace}
\usepackage{epsfig}
\usepackage{bm}
\newcommand{\be}{\begin{equation}}
\newcommand{\ee}{\end{equation}}
\newcommand{\bear}{\begin{eqnarray}}
\newcommand{\eear}{\end{eqnarray}}
\newcommand{\lomega}{\ensuremath{\lambda_\omega}}
\newcommand{\lrho}{\ensuremath{\lambda_\rho}}
\begin{document}
\title{\Large \bf Asymmetric Nuclear Matter : A variational Approach}
\author{S. Sarangi}
\affiliation{ICFAI Institute of Science \& 
Technology, Bhubaneswar-751010, India}
\author{P. K. Panda}
\affiliation{Indian Association for the Cultivation of Sciences, 
Jadavpur, Kolkata-700 032, India}
\author{S. K. Sahu}
\affiliation{Physics Department, Banki College, Banki-754008, Cuttack, India}
\author{L.Maharana}
\affiliation{Physics Department, Utkal University, Bhubaneswar-751004, India}
\email{lmaharan@iopb.res.in}
\begin{abstract}
We discuss here a self-consistent method to calculate the properties of 
the cold asymmetric nuclear matter. In this model, the nuclear matter is 
dressed with 
$s$-wave pion pairs and the nucleon-nucleon (N-N) interaction is mediated by 
these pion pairs, $\omega$ and $\rho$ mesons. The parameters of these 
interactions are calculated self-consistently to obtain the saturation 
properties like equilibrium binding energy, pressure, compressibility and 
symmetry energy. The computed equation of state is then used in the Tolman-
Oppenheimer-Volkoff (TOV) equation to study the mass and radius of a neutron 
star in the pure neutron matter limit.
\end{abstract}
\maketitle
\section {Introduction}
The search for an appropriate nuclear equation of state has been an area of 
considerable research interest because of its wide and far reaching relevance 
in heavy ion collision experiments and nuclear astrophysics. In particular, the 
studies in two obvious limits, namely, the symmetric nuclear matter (SNM) and 
the pure neutron matter (PNM) have helped constrain several properties of 
nuclear matter such as binding energy per nucleon, compressibility modulus, 
symmetry energy and its density dependence 
at nuclear saturation density $\rho_0$~\cite{Prakash97,Lattimer00,Steiner05} to varying
degrees of success. Of late, the avaliability of flow data 
from heavy ion collision experiments and phenomenological data from observation of 
compact stars have renewed the efforts to further constrain these properties and to 
explore their density and isospin content (asymmetry) variation behaviours 
\cite{Danielewicz02,Nice05,Klahn06,Piekarewicz07}. 

One of the fundamental concerns in the construction of nuclear equation of 
state is the parametrization of the nucleon-nucleon (N-N) interaction. 
Different approaches have been developed to address this problem. These 
methods can be broadly classified into three general types~\cite{Fuchs06}, 
namely, the {\it ab initio} methods, the effective field theory approaches 
and calculations based on phenomenological density functionals. The {\it ab initio} 
methods include the Brueckner-Hartree-Fock (BHF) \cite{Jaminon89,Zhou04,Baldo07} approach, 
the (relativistic) Dirac-Brueckner-Hartree-Fock (DBHF) 
\cite{Brockmann90,Li92,Jong98,GB99,Dalen07}
calculations, the Green Function Monte-Carlo (GFMC) \cite{Carlson03,Dickhoff04,Fabrocini05} 
method using the basic N-N interactions given by boson exchange 
potentials. The other approach of this type, also known as the 
variational approach, is pioneered by the Argonne Group 
\cite{Akmal97,Akmal98}. This method is also based on
basic two-body (N-N) interactions in a non-relativistic formalism with 
relativistic effects introduced successively at later stages. The effective field 
theory (EFT) approaches are based on density functional theories
\cite{Serot97,Furnstahl04} like chiral perturbation theory
\cite{Lutz00,Finelli03}. These calculations involve
a few density dependent model parameters evaluated iteratively. 
The third type of approach, namely, the calculations based on 
phenomenological density functionals include models with effective 
density dependent interactions such as Gogny or Skyrme 
forces~\cite{Bender03} and the relativistic mean field (RMF) 
models~\cite{Walecka74,Ring96,Bunta03,Liu07}. The 
parameters of these models are evaluated by carefully fitting the bulk 
properties of nuclear matter and properties of closed shell nuclei to 
experimental values. Our work presented here belongs to this class of 
approaches in the non-relativistic approximation.

The RMF models represent the N-N interactions
through the coupling of nucleons with isoscalar scalar $\sigma$ mesons, 
isoscalar vector $\omega$ mesons, isovector vector $\rho$ mesons 
and the photon quanta besides the self- and cross-interactions among 
these mesons~\cite{Bunta03}.
Nuclear equations of state have also been constructed using the quark meson
coupling model (QMC) \cite{ST} where baryons are described as systems of
non-overlapping MIT bags which interact through effective scalar and
vector mean fields, very much in the same way as in the RMF model. The QMC model
has also been applied to study the asymmetric nuclear matter at finite 
temperature \cite{panda03}.

It has been shown earlier~\cite{Mishra90,Mishra92}, that the medium and long 
range attraction effect simulted by the $\sigma$ mesons in RMF theory can 
also be produced by the $s$-wave pion pairs. This ``dressing'' of nucleons by 
pion pairs has also been applied to study the properties of 
deuteron\cite{panda92} and $^4$He \cite{panda96}. On this basis, we start with 
a nonrelativistic Hamiltonian density with $\pi N$ interaction. 
The $\omega-$repulsion and the 
isospin asymmetry part of the NN interaction are parametrized by two additional
terms representing the coupling of nucleons with the $\omega$ and the 
$\rho$ mesons respectively. The parameters of these interactions are 
then evaluated self-consistently by using the saturation properties 
like binding energy per nucleon, pressure, compressibility and the 
symmetry energy. The equation of state (EOS) of asymmetric nuclear matter is 
subsequently calculated and compared with the results of other independent 
approaches available in current literature. The EOS of pure neutron 
matter is also used to calculate the mass and radius of a neutron star.
We organize the paper as follows: In Section~II, we present the 
theoretical formalism of the asymmetric nuclear matter as outlined above.
The results are presented and discussed in Section III. Finally, in the
last section the concluding remarks are drawn indicating the future outlook 
of the model.    
\section{Formalism}
\label{Formalism}
We start with the effective pion nucleon Hamiltonian 
\begin{equation}
{\cal H}(\mathbf x)={\cal H}_N(\mathbf x)+{\cal H}_{int}(\mathbf x)+
{\cal H}_M({\mathbf x}),\label{e2}
\end{equation} 
where the free nucleon part ${\cal H}_N (\mathbf x)$ is given by 
\begin{equation}
{\cal H}_N(\mathbf x)=\psi^\dagger(\mathbf x)~\varepsilon_x~\psi(\mathbf x),
\label{e3}
\end{equation}
the free meson part ${\cal H}_M(\mathbf x)$ is defined as
\begin{equation}
{\cal H}_M(\mathbf x)={1\over 2}\left[{\dot \varphi}_i^2
+(\mbox{\boldmath $\nabla$}\varphi_i)\cdot(\mbox{\boldmath$\nabla$}\varphi_i)
+m^2\varphi_i^2\right],\label{e4}
\end{equation}
and the $\pi N$ interaction~\cite{Mishra90} is provided by
\begin{equation}
{\cal H}_{int}(\mathbf x)=\psi^\dagger(\mathbf x) \left[
-{iG\over 2 \epsilon_x }\mbox{\boldmath$\sigma$}\cdot \mathbf p ~\varphi +
{G^2\over 2 \epsilon_x }\varphi^2\right]\psi(\mathbf x).\label{e5}
\end{equation}
In equations (\ref{e3}) and (\ref{e5}), $\psi$ represents the non-relativistic 
two component spin-isospin quartet nucleon field. The single 
particle nucleon energy operator $\epsilon_x$ 
is given by $\epsilon_x=(M^2-\mbox{\boldmath $\nabla$}_x^2)^{1/2}$
with nucleon mass $M$ and the pion-nucleon coupling constant $G$. The isospin
triplet pion fields of mass $m$ are represented by $\varphi$.

We expand the pion field operator $\varphi_i(\mathbf x)$ in terms
of the creation and annihilation operators of off-mass shell pions
satisfying equal time algebra as
\begin{equation}
\varphi_i(\mathbf x)={1\over \sqrt{2 \omega_x}}
(a_i(\mathbf x)^\dagger +a_i(\mathbf x)),~~~~~~~~~~
\dot\varphi_i(\mathbf x)=i{\sqrt{\omega_x\over 2}}
(a_i(\mathbf x)^\dagger -a_i(\mathbf x)),
\label{e6}
\end{equation}
with energy $\omega_x = (m^2-\mbox{\boldmath $\nabla$}_x^2)^{1/2}$ in the 
perturbative basis. We continue to use the perturbative basis, but note 
that since we take an arbitrary number of pions in the unitary 
transformation $U$ in equation (\ref{e8}) as given later, the results would be 
nonperturbative. The expectation value of the first term of ${\cal H}_{int}(\mathbf x)$
in eq.~(\ref{e5}) vanishes and the pion pair of the second term provides the 
isoscalar scalar interaction of nucleons thereby simulating the effects 
of $\sigma$-mesons. A pion-pair creation operator given as
\begin{equation}
B^{\dag} =\frac{1}{2}\int {f}(\mathbf k)~a_i(\mathbf k)^{\dag}~a_i
(-\mathbf k)^{\dag} d \mathbf{k},\label{e7}
\end{equation}
is then constructed in momentum space with 
the ansatz function ${f}(\mathbf k)$ to be determined later. 

We then define the unitary transformation $U$ as
\begin{equation}
U=e^{(B^{\dag}-B)},
\label{e8}
\end{equation}
and note that $U$, operating on vacuum, creates an arbitrarily large number of 
scalar isospin singlet pairs of pions. The ``pion dressing'' of 
nuclear matter is then introduced through the state
\begin{equation}
|f\rangle =U|vac\rangle=e^{(B^\dagger-B)}|vac\rangle ,
\label{e9}
\end{equation}
where $U$ constitutes a Bogoliubov transformtion given by
\begin{equation}
U^\dagger~a_i(\mathbf k)U=(\cosh\ f(\mathbf k))~a_i(\mathbf k)+
(\sinh\ f(\mathbf k))~a_i(-\mathbf k)^\dagger,\label{e11}
\end{equation}

We then proceed to calculate the energy expectation values. We consider $N$ nucleons 
occupying a spherical volume of radius $R$ such that the density 
$\rho = N/({4\over 3}\pi R^3)$ remains constant as $(N,\ R)\rightarrow\infty$ 
and we ignore the surface effects. We describe the system with a density 
operator $\hat{\rho}_N$ such that its matrix elements are given by~\cite{Mishra90}
\begin{equation}
\rho _{\alpha \beta }(\mathbf x ,\mathbf y) =
Tr[\hat \rho _N ~\psi_ \beta (\mathbf y)^{\dagger}\psi_ \alpha (\mathbf x)],
\label{e13}
\end{equation}
and
\begin{equation}
Tr[\hat \rho_N \hat N]=\int \rho _{\alpha \alpha }(\mathbf x, \mathbf x) d\mathbf x
=N=\rho V.\label{e14}
\end{equation}
We obtain the free nucleon energy density 
\begin{equation}
h_f  = \langle f|Tr[\hat \rho _N {\cal H}_N(\mathbf x)]|f\rangle
= \sum_{\tau=n,p} {\gamma {k^\tau_f}^3 \over 6\pi^2}
\left (M+{3\over 10}{k_f^{\tau 2}\over M}\right ).
\label{e15}
\end{equation}
In the above equation, the spin degeneracy factor $\gamma$~=~2, the index $\tau$ 
runs over the isospin degrees of freedom $n$ and $p$ and $k^\tau_f$ represents the
Fermi momenta of the nucleons. For asymmetric nuclear matter, we define the
neutron and proton densities $\rho_n$ and $\rho_p$ respectively over the same spherical 
volume such that the nucleon density $\rho = \rho_n + \rho_p$. The Fermi momenta 
$k^\tau_f$ are related to neutron and proton densities by the relation 
$k^\tau_f = ({6\pi^2\rho_\tau / \gamma})^{1 \over 3}$. We also define the 
asymmetry parameter $y = {(\rho_n-\rho_p)/ \rho}$. It can be easily seen 
that $\rho_\tau = {\rho \over 2} (1 \pm y)$ for $\tau = n,p$ respectively.

Using the operator expansion of equation~(\ref{e6}), 
the free pion part of the Hamiltonian as given in equation~(\ref{e4}) can be 
written as
\begin{equation}
{\cal H}_M(\mathbf x)=a_i(\mathbf x)^\dagger~\omega_x~a_i(\mathbf x).
\label{e16}
\end{equation}
The free pion kinetic energy density is given by
\begin{equation}
h_k  = \langle f|{\cal H}_M (\mathbf x)|f \rangle 
= \frac{3}{(2\pi)^{3}}\int d\mathbf k ~\omega(\mathbf k)~
\mathrm{sinh}^2\ f(\mathbf k),
\label{e17}
\end{equation}
where $\omega (\mathbf k)=\sqrt{\mathbf k^2+m^2}$.
Using $\epsilon_x\simeq M$ in the nonrelativistic limit, 
the interaction energy density $h_{int}$ can be written from equation 
(\ref{e5}) as 
\begin{equation}
h_{int} = \langle f|Tr[\hat \rho_N~{\cal H}_{int}(\mathbf x)]|f \rangle
\simeq {G^2\rho\over 2 M} \langle f|:\varphi_i(\mathbf x)\varphi_i(\mathbf x):
|f \rangle.
\label{e18}
\end{equation}
Using the equations~(\ref{e8}), (\ref{e9}) and (\ref{e11}), 
we have from equation (\ref{e18}) 
\begin{equation}
h_{int}={G^2\rho\over 2M} \left({\frac{3}{(2\pi)^3}}\int 
{d\mathbf k \over \omega
(\mathbf k)} \left({\mathrm{sinh}\ 2f(\mathbf k)\over 2}+
\mathrm{sinh}^2\ f(\mathbf k)\right)\right).
\label{e19}
\end{equation}
The pion field dependent energy density terms add up to give
$h_m (=h_{k}+h_{int})$ which is to be optimized with respect to the ansatz 
function $ f(\mathbf k)$ for its evaluation. However, this ansatz function 
yields a divergent value for $h_m$. This happens because we have taken the
pions to be point like and have assumed that they can approach as near each 
other as they like, which is physically inaccurate.
Therefore, we introduce a phenomenological repulsion energy between 
the pions of a pair given by
\begin{equation}
h_m^R=\frac{3a}{(2\pi)^{3}}\int (\mathrm{sinh}^2\ f(\mathbf k))
~e^{R_\pi^2k^2}d\mathbf k,\label{e23}
\end{equation}
where the two parameters $a$ and $R_\pi$ correspond to the strength and length
scale, repectively, of the repulsion and are to be determined
self-consistently later. Thus the pion field dependent term of the total energy
density becomes $h_m=h_{k}+h_{int}+h_m^R$. Then the optimization of $h_m$ with
respect to $f(\mathbf k)$ yields  
\begin{equation}
\mathrm{tanh}\ 2f(\mathbf k)=-{G^2 \rho \over 2 M}\cdot {1\over {\omega ^2
(\mathbf k)+{G^2 \rho \over 2 M}+a \omega (\mathbf k)e^{R_\pi^2k^2}}}.
\label{e22}
\end{equation}
The expectation value of the pion field dependent parts of the 
total Hamiltonian density of eqn.~(\ref{e2}) alongwith the modification 
introduced by the phenomenological term $h_m^R$ becomes
\begin{equation}
h_m  = -{3\over 2}\frac{1}{(2\pi)^{3}} \Big( {G^2 \over{2M}} \Big)^2 
\rho \Big[ \rho_n I_n + \rho_p I_p \Big] \label{e24}
\end{equation}.
with the integrals $I_\tau$ ($\tau = n,p$) given by
\begin{equation}
I_\tau =\hspace{-0.1cm} \int_0^{k_f^\tau}{4 \pi k^2 dk\over\omega^2}
\Big[{1\over{(\omega +a e^{R_\pi^2k^2})^{1/2} (\omega +a
e^{R_\pi^2k^2}+{G^2 \rho\over M\omega})^{1/2}+(\omega +a
e^{R_\pi^2k^2})+{G^2 \rho \over 2 M\omega }}}\Big]
\label{e25}
\end{equation}
and $\omega =\omega(\mathbf k)$. 

We now introduce the energy of $\omega$ repulsion by the simple form 
\begin{equation}
h_\omega=\lomega\rho^2,\label{e26}
\end{equation}
where the parameter $\lomega$ corresponds to the strength of the interaction 
at constant density and is to be evaluated later. We note that equation~(\ref{e26}) 
can arise from a Hamiltonian density given in terms of a local potential $v_R(\mathbf x)$ as
\begin{equation}
{\cal H}_R(\mathbf x)= \psi(\mathbf x)^\dagger\psi(\mathbf x)\int v_R(\mathbf x -\mathbf y)
\psi(\mathbf y)^\dagger\psi(\mathbf y)d\mathbf y,\label{e27}
\end{equation}
where, when density is constant, we in fact have
\[ \lomega =\int v_R(\mathbf x)d\mathbf x~. \]

The isospin dependent interaction is mediated by the isovector vector $\rho$
mesons. We represent the contribution due to this interaction, in a manner 
similar to the $\omega$-meson energy, by the term
\begin{equation}
h_\rho=\lrho\rho_3^2\label{e28}
\end{equation}
where $\rho_3= (\rho_n - \rho_p)$ and the strength parameter \lrho is to be determined
as described later. Thus we finally write down the binding energy per nucleon $E_B$ of the 
cold asymmetric nuclear matter:
\begin{equation}
E_B = {\varepsilon\over\rho} - M ,
\label{e29}
\end{equation}
where $\varepsilon = (h_f + h_m + h_\omega + h_\rho)$ is the energy density. 
The expression for $\varepsilon$ contains the four model parameters $a$, 
$R_\pi$, \lomega ~and \lrho ~as introduced above. These parameters are then 
determined self-consistently through the saturation properties of nuclear 
matter. The pressure $P$, compressibility modulus $K$ and the 
symmetry energy $E_{sym}$ are given by the standard relations:
\begin{eqnarray}
P & = &\rho^2 {{\partial(\varepsilon/\rho)}\over{\partial\rho}}\label{e30} \\
K & = & 9 \rho^2 {\partial^2(\varepsilon/\rho)\over \partial\rho^2}\label{e31}\\
E_{sym} & = & \left( {1\over 2} 
{{\partial^2 (\varepsilon/\rho})\over\partial{y^2}}\right )_{y=0}. 
\label{e32}
\end{eqnarray}
The effective mass $M^\ast$ is given by
$M^\ast = M + V_s$ with $V_s=(h_{int} + h_m^R)/\rho.$

\section{Results and Discussion}
We now discuss the results obtained in our calculations and compare with 
those available in literature. The four parameters of the model are fixed by 
self-consistently solving eqs.~(\ref{e29}) through (\ref{e32}) for the respective properties 
of nuclear matter at saturation density $\rho_0$ = 0.15 fm$^{-3}$.
While pressure $P$ vanishes at saturation
density for symmetric nuclear matter (SNM), the values of binding energy per nucleon 
and symmetry energy are chosen to be $-16$~MeV and 31~MeV respectively. 
In the numerical calculations, we have used the nucleon mass $M=940$ MeV, the meson masses
$m=140$ MeV, $m_\omega=783$ MeV and $m_\rho=770$ MeV and the $\pi-N$ coupling constant 
$G^2/4\pi=14.6$. In order to ascertain the dependence of compressibility modulus on the
parameter values, we vary the $K$ value over a range 210 MeV to 280 MeV
for the symmetric nuclear matter ($y$ = 0) and evaluate the parameters. 
It may be noted that this is the range of the 
compressibility value which is under discussion in the current literature. 
For $K$ values in the range 210~MeV to 250~MeV, the program does not converge. 
The solutions begin to converge for compressibility modulus $K$ around 258 MeV. 
We choose the value $K$= 260~MeV for our calculations.
In Table~\ref{table1} we present the four free parameters of the model 
for ready reference.

\begin{table}[ht]
\caption{Parameters of the model obtained by solving the equations~(\ref{e29})-
(\ref{e32}) self consistently at saturation density.} 
\begin{tabular}{cccc} \hline \hline
a      &   $R_\pi$     &  \lomega   &   \lrho     \\
(MeV)  &    (fm)       &   (fm$^2$)   &  (fm$^2$)     \\
\hline
16.98  &   1.42 &   3.10   &   0.65 \\
\hline \hline
\end{tabular}
\label{table1}
\end{table}

\begin{figure}[ht]
\centerline{\psfig{file=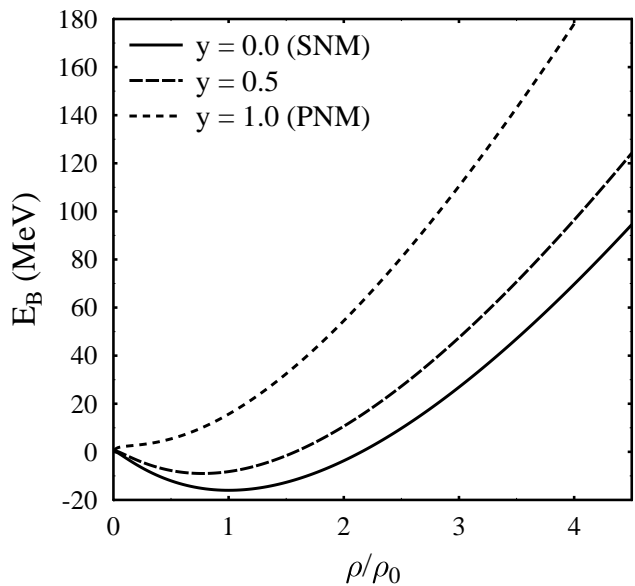,width=3.5in}}
\vspace{-2cm}
\caption{The binding energy per nucleon $E_B$ as a
function of relative nucleon density $\rho/\rho_0$ calculated for different 
values of the asymmetry parameter $y$. The values $y$ = 0.0 and 1.0 
correspond to 
symmetric nuclear matter (SNM) and pure neutron matter (PNM) respectively.}
\label{fig:eby}
\end{figure}

\begin{figure}[ht]
\centerline{\psfig{file=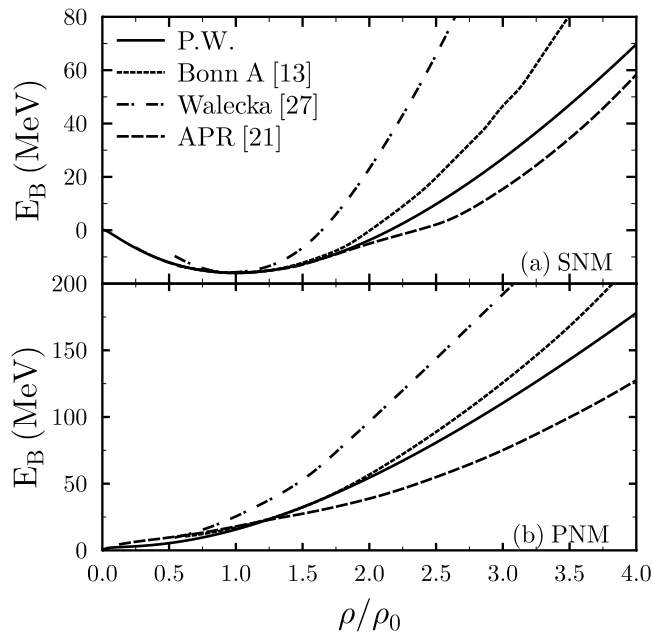,width=3.5in}}
\vspace{-1.5cm}
\caption{(a)The binding energy per nucleon $E_B$ as a
function of relative nucleon density $\rho/\rho_0$ for SNM. 
The results of present work (P.W.) are compared with the results of DBHF calculations with Bonn A 
potential ~\cite{Li92}, the variational calculations of the Argonne group 
\cite{Akmal98} and the Walecka model \cite{Walecka74}. The data for the Bonn A 
and Walecka model curves are 
taken from~\cite{Li92}. (b) Same as Fig-(2a), but for PNM.}
\label{fig:eb}
\end{figure}

For this set of parameter values the effective
mass of nucleons at saturation density is found to be ${M^\ast /M} =\ 0.81$. 
In the Fig.~\ref{fig:eby}, we present the binding energy per nucleon $E_B$ 
calculated for 
different values of the asymmetry parameter $y$ as a function of the relative
nuclear density $\rho/\rho_0$. The values $y$ = 0.0 and 1.0 
correspond to SNM and PNM respectively. As expected, the binding energy per nucleon 
$E_B$ of SNM initially decreases 
with increase in density, reaches a minimum at $\rho=\rho_0$ and then increases.
In case of PNM, the binding energy increases monotonically with increasing 
density in consistence with its well known behaviour.
In Fig.~\ref{fig:eb}(a), we compare the $E_B$ of SNM as a function of the nucleon density 
with a few representative results in the literature, namely,  
the Walecka model~\cite{Walecka74} (long-short dashed curve), the DBHF 
calculations of Li {\it et al.} with Bonn A potential 
(short-dashed curve) (data for both the models are taken from ~\cite{Li92}) and 
the variational A18 + $\delta$v + UIX* (corrected) model of Akmal {\it at al.} 
(APR) \cite{Akmal98} (long-dashed curve). While the Walecka and Bonn A 
models are relativistic, the variational model is nonrelativistic with 
relativistic effects and three body correlations introduced successively. 
Our model produces an EOS softer than that of Walecka and Bonn A, but 
stiffer than the variational calculation results of the Argonne group. 
It is well-known that the Walecka model yields a very high compressibility 
$K$. However, its improvised versions developed later with self- and cross-couplings 
of the meson fields have been able to bring down the compressibility
modulus in the ball park of 230$\pm$10 MeV \cite{Piekarewicz07}. 
Our model yields  
nuclear matter saturation properties correctly alongwith the compressibility 
of $K=260$ MeV which is resonably close to the empirical data.
In Fig.~\ref{fig:eb}(b), we plot $E_B$ as a function of the relative nucleon 
density for PNM. Similar to the SNM case, our EOS is 
softer than that of Walecka and Bonn A models, but stiffer than 
the variational model. We use this EOS to calculate the mass and radius of 
a neutron star of PNM as discussed later.

\begin{figure}[ht]
\centerline{\psfig{file=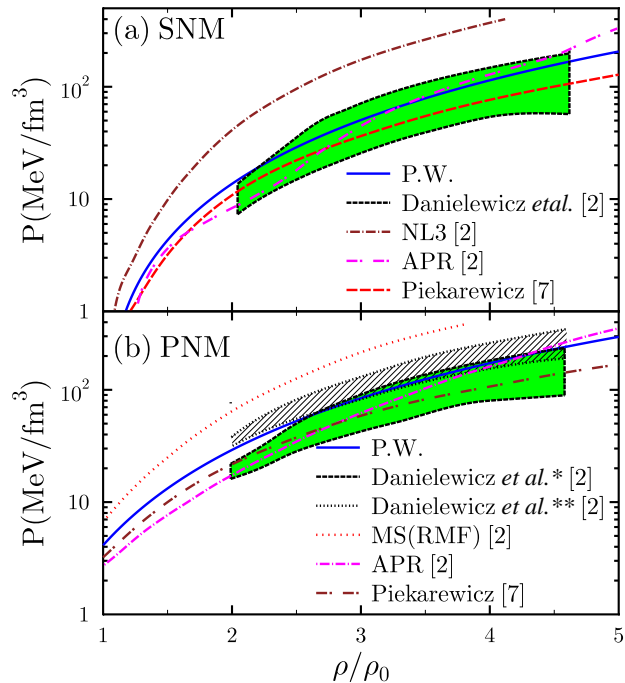,width=3.5in}}
\vspace{-3cm}
\caption{(a) The pressure as a function of relative nucleon density for SNM 
as generated by the present work (P.W.) (solid blue curve). 
The color-filled region in green corresponds 
to the bounds deduced from experimental flow data and simulations studies by
Danielewicz {\it et al.} \cite{Danielewicz02}.
The data for the curves corresponding to RMF(NL3) calculations and 
the variational calculations of Akmal {\it et al.} (APR) are taken from 
\cite{Danielewicz02}.(b) Pressure as a function of relative nucleon density 
for PNM. The shaded region and the color-filled region in green correspond 
to the bounds deduced by Danielewicz {\it et al.} using the ``stiff'' and ``soft'' 
parametrizations of Prakash {\it et al.} \cite{Prakash88}. Our EOS 
is consistent with these bounds in the cases of both SNM and PNM.}     
\label{fig:pressure}
\end{figure}

The density dependence of pressure of SNM and PNM are calculated using the 
eqn.~\ref{e30}. These results are plotted (solid blue curves) in 
Figs.~\ref{fig:pressure}(a) and (b). Recently, Danielewicz {\it et al.} 
\cite{Danielewicz02} have deduced the empirical bounds 
on the EOS in the density range of $2 ~< ~ \rho/\rho_0 ~< ~ 4.6$ by analysing the 
flow data of matter from the fireball of Au+Au heavy ion collision
experiments both for SNM and PNM. These bounds are represented by the 
color-filled and shaded regions of the two figures. These bounds rule out 
both the ``very stiff'' and the ``very soft'' classes of EOSs produced, for 
example, by some variants of RMF calculations and Fermi motion of a pure 
neutron gas~\cite{Danielewicz02}. As shown in these figures, the EOS of SNM and PNM 
generated by our model are consistent with both the bounds. 

\begin{figure}[ht]
\centerline{\psfig{file=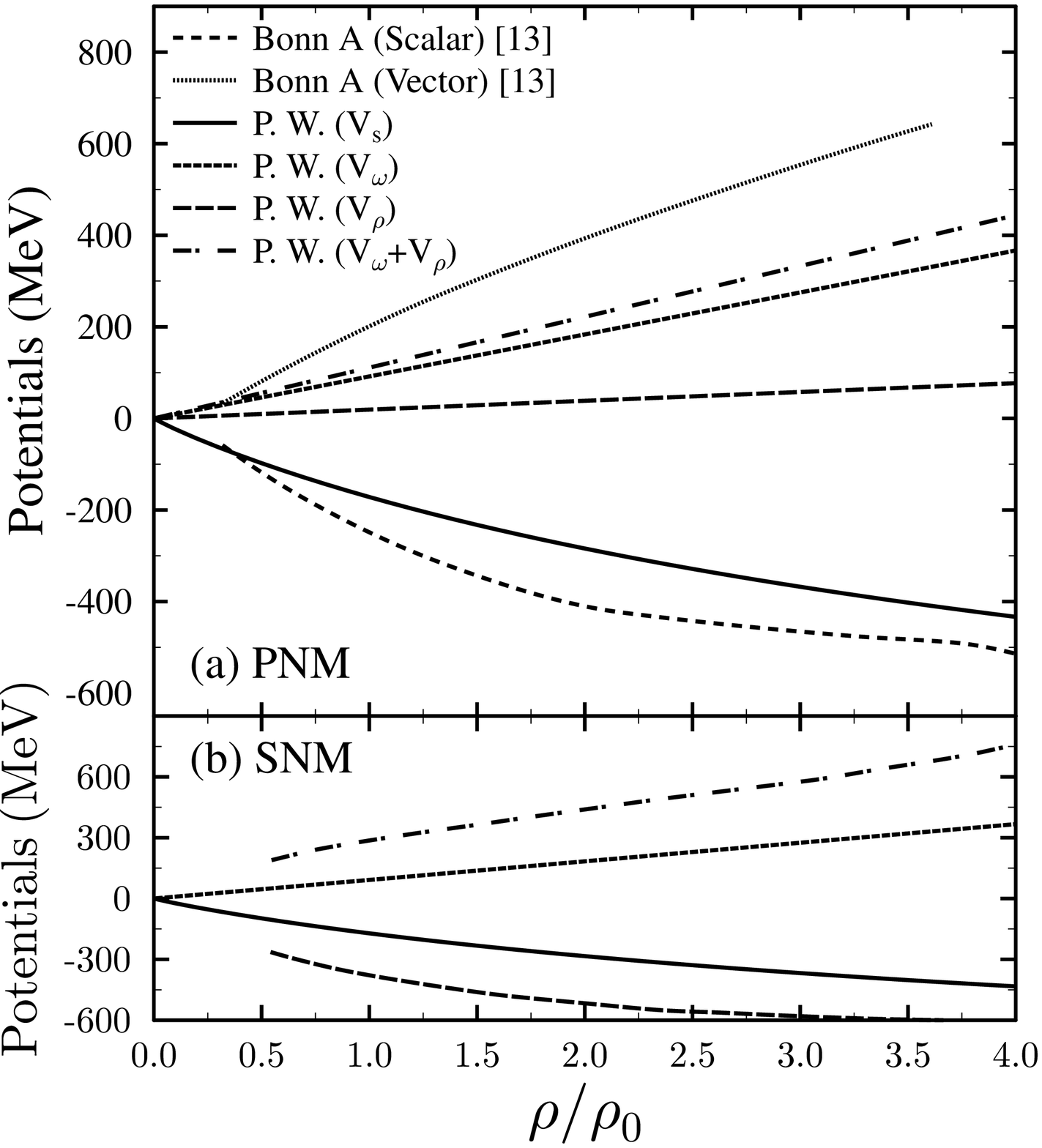,width=3.5in}}
\vspace{-8pt}
\caption{(a) The potentials $V_s$, $V_\omega$ and $V_\rho$ (as defined in the 
text) in PNM as calculated by our model are compared with
the Bonn A results of Li {\it et al.}\cite{Li92}. The contributions made by
the $\omega$-meson (short-dashed curve) and $\rho$-meson (long-dashed curve) 
mediated interactions are distinctly shown for comparison.
(b) The potentials in SNM. Because of isospin symmetry, $V_\rho$ 
(see text for definition) vanishes. Both the scalar (solid curve) and vector 
(short-dashed curve) potentials produced by our calculations are weaker in 
magnitude compared to those of Bonn A calculations.} 
\label{fig:pot}
\end{figure}

The potentials per nucleon in our model can be defined from the meson 
dependent energy terms of eqs.~(\ref{e24}), (\ref{e26}) and (\ref{e28}). 
Contribution to potential from the scalar part of the meson interaction is 
due to the pion condensates and is given by $V_s=(h_{int} + h_m^R)/\rho$ as 
defined earlier. The contribution by vector mesons has two components, namely, 
due to the $\omega$ and the $\rho$ mesons and is given by 
$V_v = V_\omega + V_\rho = (h_\omega + h_\rho)/\rho$.
In the Figs.~\ref{fig:pot}~(a) and (b), we plot $V_s$ and $V_v$ as functions of 
relative density $\rho/\rho_0$ calculated for PNM (Fig.~\ref{fig:pot}(a)) 
and for SNM (Fig.~\ref{fig:pot}(b)) respectively.  
The magnitudes of the potentials calculated by our model are weaker compared 
to those produced by DBHF calculations with Bonn A interaction~\cite{Li92} 
as shown in both the panels of Fig.~\ref{fig:pot}. In Fig.~\ref{fig:pot}(a), 
we show the contributions to the repulsive vector potential due to $\omega$ 
mesons (short-dashed curve), $\rho$ mesons (long-dashed curve) and their 
combined contribution (long-short-dashed curve). The contribution
due to $\rho$ mesons rises linearly at a slow rate and has a low contribution 
at saturation density. This indicates that major contribution to the 
short-range repulsion part of nuclear force is from $\omega$ meson interaction.
\begin{figure}[ht]
\centerline{\psfig{file=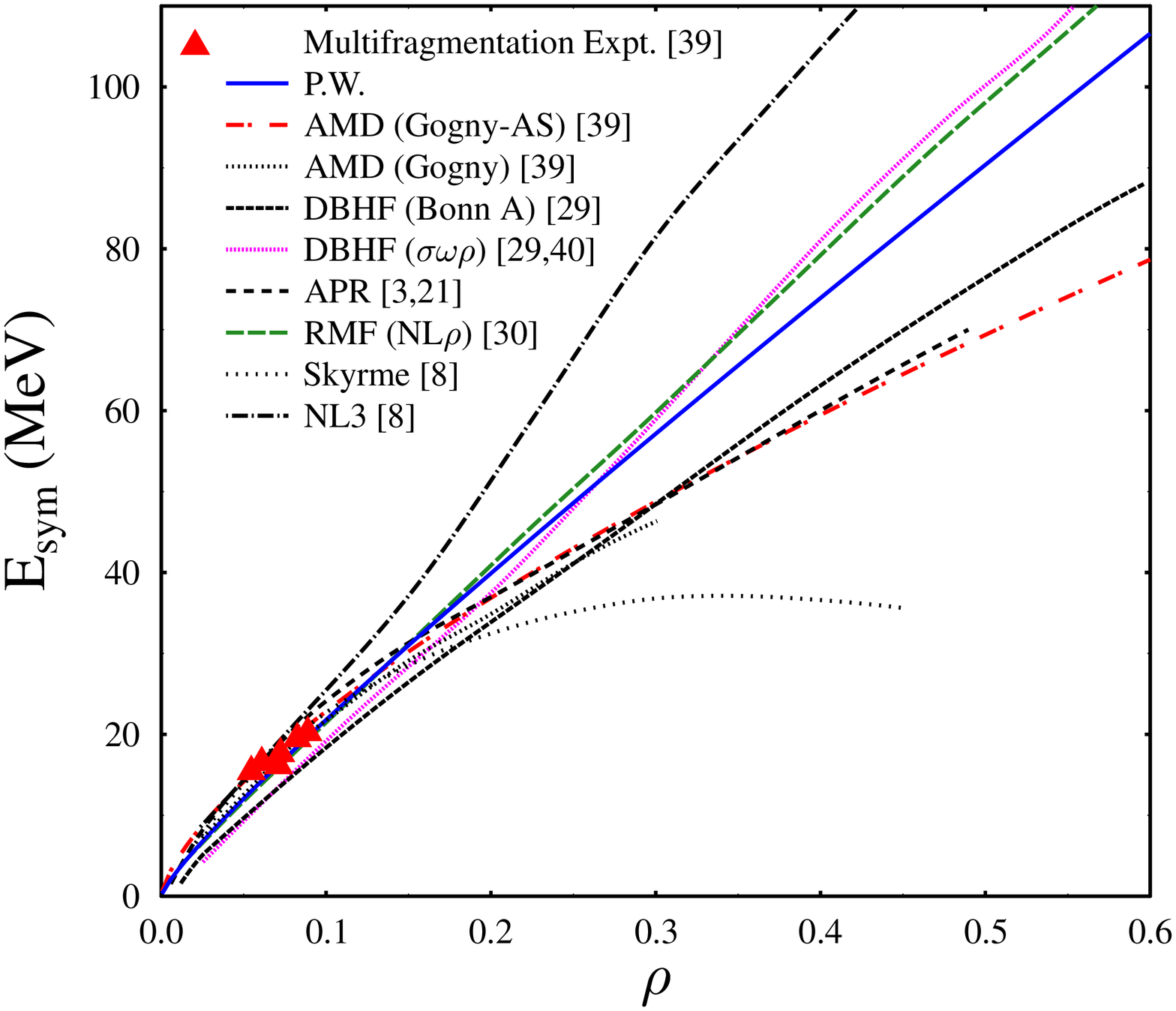,width=4.0in}}
\vspace{-1.5cm}
\caption{Symmetry energy $E_{sym}$ calculated from the EOS (as in 
Eq.~\ref{e32}) (P.W.) (solid blue line) is plotted as a function of density along 
with results of other groups. The data for experimental points and the results 
of the antisymmetrized molecular dynamics (AMD) simulations with Gogny-AS and 
Gogny interactions are taken from Shetty~{\it et al} \cite{Shetty07}, 
DBHF (Bonn A) results are taken from \cite{Bunta03}, RMF (NL$\rho$) data are 
from \cite{Liu07}, the variational model of Akmal {\it et al.} 
(APR)~\cite{Akmal98} results are from \cite{Steiner05}, DBHF ($\sigma \omega 
\rho$) model of Huber {\it et al.}~\cite{Huber} data are from \cite{Bunta03}, 
the Skyrme amd NL3 results are 
from~\cite{Fuchs06}. Our result shows consistency with those of other groups 
and corroborates the moderately ``stiff'' dependence 
of $E_{sym}$ as advocated by Shetty \it{et al.}~\cite{Shetty07}.}
\label{fig:esym}
\end{figure}                                     

Knowledge of density dependence of symmetry energy is expected to play a key 
role in understanding the structure and properties of neutron-rich nuclei 
and neutron stars at densities above and below the
saturation density. Therefore this problem has been receiving 
considerable attention of late. Several theoretical and experimental 
investigations addressing this problem have been reported~
(\cite{Steiner05,Fuchs06,Shetty07} and references therein). While the results 
of independent studies show reasonable consistency at
sub-saturation densities $\rho \leq \rho_0$, they are at wide variance 
with each other at supra-saturation densities $\rho > \rho_0$. This wide 
variation has given rise to the so-called classification of ``soft'' and 
``stiff'' dependence of symmetry energy on density~ \cite{Stone03,Shetty07}.

Fig.~\ref{fig:esym} shows a representation of the spectrum 
of such results alongwith the results of the present work (solid blue curve). 
While the Gogny and Skyrme forces (dark rib-dotted and dotted curves respectively with 
data taken from \cite{Fuchs06,Shetty07}) produce ``soft'' dependence 
on one end, the NL3 force (dot-dashed curve with data taken from~\cite{Fuchs06}) 
produces a very ``stiff'' dependence on the other end. The analysis of 
experimental and simulation studies of intermediate energy heavy-ion reactions 
as reported by Shetty {\it et al.}~\cite{Shetty07} (red triangles and long-short-dashed 
red curve repectively), results of DBHF calculations of Li {\it et al.} and 
Huber {\it et al.}~\cite{Li92,Bunta03,Huber} (rib-dashed and magenta 
ribbed curve), variational model~\cite{Steiner05,Akmal98} (short-dashed curve), RMF 
calculations with nonlinear Walecka model including $\rho$ mesons by Liu 
{\it et al.}\cite{Liu07} (long-dashed green curve) as shown in Fig.~\ref{fig:esym} 
suggest ``stiff'' dependence with various degrees of stiffness. The experimental 
results (represented by the red triangles with data taken from 
Shetty {\it et al.}~\cite{Shetty07}) are derived 
from the isoscaling parameter $\alpha$ which, in turn, is obtained from relative 
isotopic yields due to multifragmentation of excited nuclei produced by bombarding 
beams of $^{58}$Fe and $^{58}$Ni on $^{58}$Fe and $^{58}$Ni targets.
Shetty {\it et al.} have shown that the results of multifragmentation simulation 
studies carried out with Antisymmetrized Molecular Dynamics (AMD) model 
using Gogny-AS interaction and Statistical Multifragmentation Model (SMM) 
are consistent with the above-mentioned experimental results and suggest 
(as shown by the red long-short-dashed curve) a moderately stiff dependence of 
the symmetry energy on density. Our results (represented by the solid blue curve) 
calculated using eqn.~(\ref{e32}) are consistent with these results 
at subsaturation densities but are stiffer at supra-saturation densities. More
observational or experimental information is required to be built into our model
to further constrain the symmetry energy at higher densities. 
In Fig.\ref{fig:esym}, the curve due to Huber~{\it et al.}~
\cite{Huber} (with data taken from \cite{Bunta03}) 
correspond to their DBHF `HD' model calculations which involves 
only the $\sigma$, $\omega$ and $\rho$ mesons. Similarly the long-dashed green 
curve due to Liu~{\it et al.}~\cite{Liu07} is from the basic non-linear 
Walecka model with $\sigma$, $\omega$ and $\rho$ mesons. Our formalism is 
the closest to these two models with the exception that in our model the 
effect of $\sigma$ mesons is simulated by the $\pi$ meson condensates. It is 
also noteworthy that our results are consistent with these results for 
densities upto $2\rho_0$. 

The wide variation of density dependence of symmetry energy at supra-saturation densities 
has given rise to the need of constraining it. As discussed by Shetty {\it et al}
\cite{Shetty07}, a general functional form $E_{sym} = E^0_{sym} (\rho / \rho_0)^\gamma$ 
has emerged. Studies by various groups have produced the fits with 
$E^0_{sym} \sim 31-33$~MeV and $\gamma \sim 0.55-1.05$. A similar parametrization 
of the $E_{sym}$ produced by our EOS with $E^0_{sym} = 31$ MeV yields the exponent 
parameter $\gamma$ = 0.85.

We next use the equation of state for PNM derived by our model in the 
Tolman-Oppenheimer-Volkoff (TOV) equation to calculate the mass and radius of 
a PNM neutron star. The mass and radius of the star are found to be $2.25M_\odot$ 
and 11.7 km respectively. 
\section{Conclusion}
In this work we have presented a quantum mechanical nonperturbative formalism 
to study cold asymmetric nuclear matter using a variational method. 
The system is assumed to be a collection of nucleons interacting via exchange 
of $\pi$ pairs, $\omega$ and $\rho$ mesons. 
The equation of state (EOS) for different values of asymmetry 
parameter is derived from the dynamics of the interacting system 
in a self-consistent manner. This formalism yields results similar 
to those of the {\it ab initio} DBHF models, variational models and the RMF 
models without invoking the $\sigma$ mesons. The compressibility modulus 
and effective mass are found to be $K$ = 260~MeV and $M^*/M$ = 0.81 
respectively. The symmetry energy calculated from the EOS  suggests a moderately
``stiff'' dependence at supra-saturation densities 
and corroborates the recent arguments of Shetty {et al.}~\cite{Shetty07}. 
A parametrization of the density dependence of symmetry energy of the form 
$E_{sym} = E^0_{sym} (\rho / \rho_0)^\gamma$ with the symmetry energy 
$E^0_{sym}$ at saturation density being 31 MeV produces $\gamma = 0.85$. 
The EOS of pure neutron matter (PNM) derived by the formalism yields the 
mass and radius of a PNM neutron star 
to be $2.25M_\odot$ and 11.7 km respectively. 
\section{Acknowledgements}
P.K.P would like to acknowledge Julian Schwinger foundation for financial 
support. P.K.P wishes to thank Professor F.B. Malik and Professor 
Virulh Sa-yakanit for inviting the CMT31 workshop. The authors are also thankful
to Professor S.P. Misra for many useful discussions.

\end{document}